\documentclass[review,12pt]{elsarticle}

\usepackage[T1]{fontenc}
\usepackage{lmodern}
\usepackage{pbox}
\usepackage{array}
\usepackage{makecell}
\usepackage{multirow}
\usepackage{dsfont}
\usepackage{url}
\usepackage{subfig}
\usepackage{amsmath,amsfonts,amssymb, amsthm,bm}
\usepackage{graphicx}
\usepackage{float}
\usepackage{booktabs,tabularx}

\usepackage{stfloats}
\usepackage[toc,page]{appendix}
\usepackage{threeparttable}
\usepackage{microtype}
\usepackage{chngcntr}
\usepackage{subfig}
\usepackage{relsize}

\usepackage{pbox}
\usepackage{color,soul}
\usepackage{array}

\usepackage{lineno}

\begin{document}
	
	\begin{frontmatter}

\title{Modelling the Hourly Consumption of Electricity during Period of Power Crisis}
\author[label1,label2]
{Samuel Asante Gyamerah}
\address[label1]{Department of Statistics and Actuarial Science, Kwame Nkrumah University of Science and Technology, Kumasi-Ghana}
\address[label2]{Laboratory for Interdisciplinary Statistical Analysis -- Kwame Nkrumah University of Science and Technology (KNUST-LISA), Kumasi-Ghana}
\author[label1]{Henry Ofoe Agbi-Kaiser}
\author[label1]{Keziah Ewura Adjoa Amankwah}
\author[label1]{Patience Anipa}
\author[label1]{Bright Arafat Bello}
\tnotetext[label1]{correspondence:\,\, saasgyam@gmail.com}

\begin{abstract}
	In this paper, we capture the dynamic behaviour of hourly consumption of electricity during the period of power crisis (``dumsor'' period) in Ghana using two-state Markov switching autoregressive (MS-AR) and autoregressive (AR) models. Hourly data between the periods of January 1, 2014, and December 31, 2014 was obtained from the Ghana Grid company and used for the study. Using different information criteria, the MS(2)-AR(4) is selected as the optimal model to describe the dynamic behaviour of electricity consumption during periods of power crisis in Ghana. The parameters of the MS(2)-AR(4) model are then estimated using the expectation-maximization algorithm. From the results, the likelihood of staying under a low electricity consumption regime is estimated to be 87\%. The expected duration for a low electricity consumption regime is 7.8 hours daily, and the high electricity consumption regime is expected to last 2.3 hours daily. The proposed model is robust as compared to the autoregressive model because it effectively captures the dynamics of electricity demand over time through the peaks and significant fluctuations in consumption patterns. Similarly, the model can identify distinct regime changes linked to electricity consumption during periods of power crises. 
\end{abstract}


\end{frontmatter}

\section{Introduction}
\noindent
While energy supply in Ghana has been relatively stable in recent years, the country has been beset by energy supply issues in the past, which have had a substantial impact on the country's economic status \cite{kumi2017electricity}. This is due to either low levels of water in the dams or technical malfunctions of equipment caused by high amounts of water. However, the issue of electricity generation and delivery does not appear to be limited to Ghana; it appears to be a concern in many developing countries, including all members of the Economic Community of West African States (ECOWAS). In Ghana, electricity consumption is constantly plagued with power outages and fluctuations popularly known as ``DumSor''. According to \cite{adeoye2019modelling}, the Economic Community of West African States (ECOWAS) aims to achieve 100\% electrification in all member countries by 2030, and that in order to do so, energy generation capacities must be greatly increased. Ghana's electricity sector is in a period of transition and its electricity consumption growth rate for the past five years has been increasing by 10.3\% annually, with supply falling short of expectations, \cite{energy2021commission}. Projections were made in line with ECOWAS' aim to make electricity accessible to all Ghanaians by 2020 \cite{eshun2016review}. But as at the end of 2019, the access rate was 83.5\% and with a growth rate of 3.13\%, the country missed the target by 14\%. The actual challenge is ensuring that this aim is met and, more importantly, that supply is both consistent and enough. Ghana, as at 2021, still experiences nationwide blackouts like on Sunday, March 7th, 2021 and load shedding exercises like the Volta and Oti Region load shedding operation from Thursday, March 18th, to Monday, March 21st, 2021. In addition, the annual increase in electricity demand has caught up with available generation, leaving very little spare capacity to deal with a system outage.
\\
[2mm]
Different techniques for modelling hourly electricity consumption have been studied in the literature. For instance, \cite{gabreyohannes2010nonlinear} used the self-exciting threshold autoregressive (SETAR) model and the smooth transition regression (STR) model to model, analyze, and forecast the residential electricity consumption in Ethiopia. The study showed that the SETAR model was more effective than the STR model. \cite{nwulu2012modelling} used a back propagation neural network to model electricity consumption. In a study conducted by \cite{pielow2012modeling}, a regression model was used to generate hourly electricity consumption over the course of a year for the commercial and industrial sectors of three US cities. The model was built using hourly datasets from the previous four years. \cite{nafidi2016modelling} considered a new extension of the stochastic Gamma diffusion process by introducing time functions to model electric power consumption during a period of economic crisis. \cite{palacios2018stochastic} employed bottom-up stochastic models to simulate high-resolution heating and cooling electricity consumption profiles. In a recent study, \cite{dalkani2021modelling} used the Markov process technique, feature selection, and clustering to model electricity consumption forecasting for Bushehr-Iran Power Distribution Company.  \cite{perez2022model} used principal component analysis to model the monthly electricity consumption of public sanitary buildings using climatological variables. 
\\
[2mm]
Globally, household electricity consumption is affected by factors like seasons, the number of occupants living in the house, the income of the household, and time of day. People also consume electricity in different ways and with different degrees of urgency. The variation in Ghana's household electricity consumption is greatly affected by the time of day, where low consumption is recorded during the early hours of the day and high consumption is usually in the evening and nighttime since most people are employed during the day. Understanding how variation in residential electricity demand affects consumers and businesses alike is crucial for sustainability programs. Typically, time series models are commonly used to evaluate the transient behaviour of variables like power demand, exchange rates, and temperature, among others. The mixed autoregressive moving averages (ARMA), moving average (MA), and autoregressive (AR) models are the most popular and often used models. Although these models are highly effective in a variety of applications, they are incapable of representing certain nonlinear dynamic patterns, such as asymmetry, amplitude dependency, and volatility clustering. For example, electricity consumption normally fluctuates during the time of the day, with high and low consumption depending on the time of the day (see Figure \ref{fig:one_month}). With this kind of data, it would be inefficient to think that a single linear model could explain all of these different behaviors. Markov switching models (MSM) are a type of nonlinear time series model that has become popular for explaining how different time series regimes behave. 
\\
[2mm]
According to \cite{hamilton1989new}, the Markov switching model consists of a set of models that may characterize time series behavioural patterns under various regimes. The switching technique is regulated by a latent state variable that assumes the process of a ``first-order Markov chain,'' which is a distinctive aspect of the Markov switching model. For instance, \cite{bierbrauer2004modeling} modeled electricity spot prices using regime-switching models. 
\cite{adom2013modelling} used full-period, pre-reform, and post-reform sample times to model the demand for electricity in Ghana based on the effects of policy regime switching. The study found that technology changes were energy-saving during the pre-reform period, while technology changes were energy-consuming during the post-reform period. Using the entire sample period, electricity demand in the long run is highly influenced by GDP and industry efficiency. \cite{gyamerah2018regime} developed a time-dependent two-state Markov Regime Switching (MRS) model to capture the hourly spot price of electricity. The constructed model was efficient as it was able to capture the main characteristics exhibited in the hourly electricity spot price. To model the dynamics of temperature for weather derivatives, \cite{gyamerah2018regime} used a time-changing mean-reversion Levy regime-switching model to capture both normal and extreme temperature fluctuations. \cite{cardenas2015markov} described ``El Nino Southern Oscillation (ENSO)'' patterns using a two-state Markov regime-switching framework. They discovered that the behavioral patterns of both research phases (El Nino and La Nina events) are diametrically opposed and distinct. The findings of the study also show that, while the Box-Jenkins approach produces a decent representation of the time series under review, it is unable to capture some nonlinearities that arise as a result of the presence of changing regimes. Furthermore, it was shown that the presence of weather cycles necessitates the origin of non-linear factors in many climatic time series for describing climatic variables. The model was able to reflect the index's properties across time. This study therefore employs a Markov regime switching model to model electricity consumption since its switching mechanism can capture the distinct behaviour of electricity consumption. Specifically, the objective of this paper is: (1) to model the consumption of electricity during periods of power crisis using a Markov-regime switching model and (2) to estimate the expected duration of each regime for the selected period.
\\
The paper is organized as follows. The Markov switching autoregressive model and parameter estimate of the model using the expectation maximization algorithm are presented in Section 2. In section 3, the data description and analysis are presented.  The results and analysis are presented in Section 4, whilst the conclusion and policy recommendations are presented in Section 5.

\section{Markov Switching Autoregressive (MS-AR) Model} 
\noindent
The MS-AR model's point of reference is the Markov chain. An approach called the Markov chain explains how the likelihood of the occurrence of an event depends solely on the attained state of a prior event. The constructed Markov switching model is based on the studies of \cite{hamilton1989new} and \cite{hamilton1994state}.

\subsection{Markov Switching (MS) Model}
\noindent
An MS model is the name given to a combination of discrete-time stochastic processes, one of which is unobserved(latent) and the other of which is observed. The dynamics of the process that is not being observed affects that of the observed. As a result, the series of observations can be used to rebuild the distribution of the unobserved process \cite{kuan2002lecture}.
Given that the Markov chain is conditionally autoregressive, the unobserved process is a Markov chain with finite-state defined as $s_t$, whereas the observable process is defined as $y_t$. 
\\
Let $s_t$ denote an unobservable state variable assuming the value one or two. A simple switching model for the variable $y_t$ involves two AR specifications:
\begin{equation}\label{eq:MS}
y_t = \left\{
\begin{array}{lr}
\alpha_0+\beta y_{t-1}+\epsilon_t, & s_t = 1\\
\alpha_0+\alpha_{1}+\beta y_{t-1}+\epsilon_t, & s_t = 2
\end{array}
\right.
\end{equation}
where $|\beta|<1$ and $\epsilon_t\thicksim i.i.d.N(0,\sigma_{\epsilon}^2)$. When $s_t = 0$, $y_t$ is a stationary AR (1) process with mean $\alpha_0 / (1-\beta)$ and when $s_t$ changes from 1 to 2, $y_t$ switches to another stationary AR(1) process with mean $(\alpha_0 + \alpha_1 )/(1-\beta)$. Based on the value of the state variable $s_t$, this model accepts two dynamic structures at various levels if and only if $\alpha_1\neq1$. $s_t$ controls the switching between these two states which are two distinct distributions with different means coming together to form $y_t$.

\subsection{Autoregressive (AR) Models}
\noindent
An Autoregressive model is a specific type of regression model for time series where a value from the series is regressed on prior value(s) of that same series. For example,
\begin{equation}\label{eq:AR1}
y_t = \alpha_0+\alpha_{1}y_{t-1}+\epsilon_t
\end{equation}
In model (\ref{eq:AR1}) response variable from the preceding time period, $y_{t-1}$, has been transformed into a predictor. Under the usual assumptions, the error terms are $\epsilon_t\thicksim i.i.d.N(0,\sigma_{\epsilon}^2)$.
\\
[2mm]
The number of immediate past values used in predicting the current time value determines the order of the AR model. A first-order autoregressive Model abbreviated as AR (1) is given by (\ref{eq:AR1}). \\
[2mm]
Using two prior values of the time series as predictors in the model changes the model from AR (1) to AR (2), a second-order autoregressive Model which can be written as;
\begin{equation}\label{eq:AR2}
y_t = \alpha_0+\alpha_{1}y_{t-1}+\alpha_{2}y_{t-2}+\epsilon_t
\end{equation}
A generalized autoregressive model of the $p$-th-order abbreviated as $AR (p)$ is obtained by incorporating prior values of the series at times $t-1, t-2,\dots,t-p$ as predictors into a multiple linear regression model to predict the current time value. This model may be expressed as follows:
\begin{equation}\label{eq:ARp}
y_t = \alpha_0+\alpha_{1}y_{t-1}+\alpha_{2}y_{t-2}+\dots+\alpha_{p}y_{t-p}+\epsilon_t
\end{equation}
The model given by (\ref{eq:ARp}), predicts $y_t$ based on the prior time values to time $t-p$.

\subsection{Markov Switching Autoregressive (MS-AR) Models}
\noindent
The MS-AR model requires a probabilistic process to govern the transition of $y_t$ from one regime (state) to another. In our case, a two-state Markov chain is used. This chain assumes that the probability of being in state $s_t$ is solely determined by the previous value of state $s_{t-1}$.
\begin{equation}\label{eq:MC}
P_{r}(s_t = j|s_{t-1} = i, s_{t-2} = k, ...) = P_{r}(s_t = j|s_{t-1} = i) = p_{ij}
\end{equation}
where the transition probability $p_{ij}$ which depicts the possibility of moving to state $j$ following that the process is in state $i$ satisfies equation (\ref{eq:TPS})
\begin{equation}\label{eq:TPS}
\mathlarger{\sum}_{j=1}^{K}\,\,p_{ij} = 1
\end{equation}
The shift between regimes in the MS-AR models is governed by an unobserved Markov chain which is based solely on the observed time series. Model (\ref{eq:MSARM}) summarized the general form of the MS-AR model
\begin{equation}\label{eq:MSARM}
(y_t-\mu_{s_t}) = \mathlarger{\sum}_{i=1}^{K}\,\,\beta_{i}(y_{t-i}-\mu_{s_{t-i}})+\epsilon_t
\end{equation}
where the relevant observations are given by $y_t$, the coefficients are $\beta_i, i = 1,\dots, K$, $s_t$ is the regime at time $t$, $\mu_{s_t}$ is a constant that depends on the regime $s_t$, and the distribution of $\epsilon_t$ is Gaussian $N(0,\sigma^2)$.\\
The conditional likelihood density model (Equation \ref{eq:CLM1}) can be used to depict the likelihood density model of $y_t$ given the stochastic variable $s_t$ for $j=1,2,\dots,K$.
\begin{equation}\label{eq:CLM1}
f(y_t|s_{t}=j;\mu_{j},\sigma_{j}^2) = \dfrac{1}{\sqrt{2\pi\sigma_{j}^2}}exp\left(\dfrac{-(y_t-\mu_j)^2}{2\sigma_{j}^2}\right)
\end{equation}
Certain stochastic distributions determine the hidden or latent regime $s_t$. Equation (\ref{eq:PROB1}) shows how the model is represented by $\pi_j$ for the marginal likelihood of $s_t$ and $j=1,2,\dots,K$.
\begin{equation}\label{eq:PROB1}
\pi_j=Pr(s_t=j;\theta)
\end{equation}
$\theta$ includes the likelihood of $\pi_j$, where $\theta = (\mu_1,\mu_2,\dots,\mu_K,\sigma_{1}^2,\sigma_{2}^2,\dots,\sigma_{K}^2,\pi_1,\pi_2,\dots,\\\pi_K)$.

\subsection{Parameter Estimation of MS-AR model}
\noindent 
The expectation maximization (EM) approach is used to estimate the model's parameters. In doing so, a preliminary estimate of the probability density function is made. This function is then used to derive the log-likelihood function. This process of estimation can be divided into two main steps.

\subsubsection{Derivation  of joint density function of $y_t$, $s_t$ and $s_{t-1}$}
\noindent 
The first step is to consider the joint density of $y_t$, $s_t$, and $s_{t-1}$ conditioned on $psi_{t-1}$ information. To achieve this, we must derive the density function of $y_t$. This is produced from a fundamental analysis of a two-state Markov switching model with a first order autoregressive term using the the MS-AR model in (\ref{eq:MSARM}) and the transition probability condition in (\ref{eq:TPS}) as shown in model equation (\ref{eq:pdf}).
\begin{equation}\label{eq:pdf}
f(y_t|s_{t},s_{t-1},\psi_{t-1};\theta) = \dfrac{1}{\sqrt{2\pi\sigma_{s_t}^2}}exp\left(\dfrac{((y_t-\mu_{s_t})-\beta_1(y_{t-1}-\mu_{s_{t-1}}))^2}{2\sigma_{s_t}^2}\right)
\end{equation}
where $\psi_{t-1}$ refers to information up to time $t-1$ and $\theta=(\mu_{1},\mu_{2},\sigma^2,\beta_{1})$ is the parameter space of the MS(2)-AR(1) model.
\\
[2mm]
Then the joint density of $y_t$, $s_t$ and $s_{t-1}$ conditioned on $\psi_{t-1}$ is:
\begin{equation}\label{eq:joint}
\begin{split}
&f(y_t,s_{t},s_{t-1}|\psi_{t-1};\theta)=f(y_t|s_{t},s_{t-1},\psi_{t-1};\theta)\times Pr(s_t,s_{t-1}|\psi_{t-1};\theta)
\end{split}
\end{equation}
where $f(y_t|s_{t},s_{t-1},\psi_{t-1};\theta)$ is given by (\ref{eq:pdf}).

\subsubsection{Aggregation of Densities Over Regime}
\noindent
Aggregating the joint density in equation (\ref{eq:joint}) over all possible values for both $s_t$ and $s_{t-1}$ yields the marginal density:
\begin{equation}\label{eq:marginal_density}
\begin{split}
&f(y_t|\psi_{t-1};\theta)\\
&= \mathlarger{\sum}_{s_t=1}^{K}\,\,\mathlarger{\sum}_{s_{t-1}=1}^{K} f(y_t,s_{t},s_{t-1}|\psi_{t-1};\theta)\\
&=\mathlarger{\sum}_{s_t=1}^{K}\,\,\mathlarger{\sum}_{s_{t-1}=1}^{K} f(y_t|s_{t},s_{t-1},\psi_{t-1};\theta)\times Pr(s_t,s_{t-1}|\psi_{t-1};\theta)
\end{split}
\end{equation}
Then the log-likelihood function is given by:
\begin{equation}\label{eq:log_likelihood}
\begin{split}
\ln L(\theta) =  \mathlarger{\sum}_{t=1}^{T}\,\,\ln\left\{\mathlarger{\sum}_{s_t=1}^{K}\,\,\mathlarger{\sum}_{s_{t-1}=1}^{K} f(y_t|s_{t},s_{t-1},\psi_{t-1};\theta)\times Pr(s_t,s_{t-1}|\psi_{t-1};\theta)\right\}
\end{split}
\end{equation}
To complete the above procedure, we still need to deal with the challenges in the calculation of  $Pr(s_t,s_{t-1}|\psi_{t-1};\theta)$. The filtering and smoothing algorithms are used to assess the probability that an observation will follow a particular pattern from one observation $t-1$ to the next $t$. 
\\
[2mm]
The filtering algorithm is illustrated with the two steps below:\\
STEP 1: At the $t$-th iteration or beginning of time $t$, $Pr(S_{t-1}=i|\psi_{t-1};\theta),
,i = 1,2,\dots,K$ is already know. Then the weighting terms $Pr(s_t=j,s_{t-1}=i|\psi_{t-1};\theta),\,i = j = 1,2,\dots,K$ can be calculated as
\begin{equation}\label{eq:PE3}
\begin{split}
&Pr(s_t=j,s_{t-1}=i|\psi_{t-1};\theta)=Pr(s_t=j|s_{t-1}=i|;\theta)\times Pr(s_{t-1}=i|\psi_{t-1};\theta)\\
\end{split}
\end{equation}
where $Pr(s_t=j|s_{t-1}=i|;\theta),\,i = j = 1,2,\dots,K$ are the transition probabilities.
\\
[2mm]
STEP 2: At the end of the $t-th$ iteration or the end time $t$, $y_t$ will be observed. We will then update the probability terms as follows:
\begin{equation}\label{eq:PE4}
\begin{split}
&Pr(s_t=j,s_{t-1}=i|\psi_{t-1},y_t;\theta)\\
&=\dfrac{f(y_t,s_{t}=j,s_{t-1}=i|\psi_{t-1};\theta)}{f(y_t|\psi_{t-1};\theta)}\\
&=\dfrac{f(y_t|s_{t}=j,s_{t-1}=i,\psi_{t-1};\theta)\times Pr(s_t=j,s_{t-1}=i|\psi_{t-1};\theta)}{\mathlarger{\sum}_{s_t=1}^{K}\,\,\mathlarger{\sum}_{s_{t-1}=1}^{K} f(y_t|s_{t},s_{t-1},\psi_{t-1};\theta)\times Pr(s_t,s_{t-1}|\psi_{t-1};\theta)}
\end{split}
\end{equation}
with
\begin{equation}\label{eq:PE5}
\begin{split}
Pr(s_t=j|\psi_{t-1},y_t;\theta)=\mathlarger{\sum}_{i=1}^{K}Pr(s_t=j,s_{t-1}=i|\psi_{t-1},y_t;\theta)\\
\end{split}
\end{equation}
Iterating the above two steps for $t=1,2,\dots,T$ provides us with the appropriate weighting terms in (\ref{eq:marginal_density}).
\\
[2mm]
The smoothing algorithm on the other hand uses all the observation in the sample. This gives $Pr(s_t=j|\psi_{T};\theta),\,t=1,2,\dots,T,$ which is the smoothed probability as opposed to $Pr(s_t=j|\psi_{t-1},y_t;\theta)=Pr(s_t=j|\psi_{t};\theta),\,t=1,2,\dots,T,$ which is the filtered probability in (\ref{eq:PE5}), \cite{kim1994dynamic}.\\
The smoothing algorithm as illustrated below considers the derivation of the joint probability that $s_t=j$ and $s_{t+1} = k$ is based on the full observation:
\begin{equation}\label{eq:PE6}
\begin{split}
&Pr(s_t=j, s_{t+1} = k|\psi_{T};\theta)\\
&=Pr(s_{t+1} = k|\psi_{T};\theta)\times Pr(s_t=j, s_{t+1} = k|\psi_{t};\theta)\\
&=\dfrac{Pr(s_{t+1} = k|\psi_{T};\theta)\times Pr(s_t=j, s_{t+1} = k|\psi_{t};\theta)}{Pr(s_{t+1} = k|\psi_{t};\theta)}\\
&=\dfrac{Pr(s_{t+1} = k|\psi_{T};\theta)\times Pr(s_t=j|\psi_{t};\theta)\times Pr(s_{t+1} = k|s_t=j;\theta)}{Pr(s_{t+1} = k|\psi_{t};\theta)}\\
\end{split}
\end{equation}
and
\begin{equation}\label{eq:PE7}
\begin{split}
Pr(s_t=j|\psi_{T};\theta)=\mathlarger{\sum}_{k=1}^{K}Pr(s_t=j,s_{t+1}=k|\psi_{T};\theta)\\
\end{split}
\end{equation}
The basic filter's last iteration produces $Pr(s_{T}|\psi_{T})$ which would be use in the above iteration for $t=T-1,T-2,\dots,1$ to generate the smoothed probabilities, $Pr(s_{t}|\psi_{T}),\,\,t=T-1,T-2,\dots,1.$
\\
[2mm]
There are two limiting models, $\pi1$ and $\pi2$, for a two regime MS-AR  model. As stated in model (\ref{eq:lagrange}), the maximum likelihood in the Lagrange model is estimated using the log-likelihood approach.
\begin{equation}\label{eq:lagrange}
\begin{split}
J(\theta)=\ln{L(\theta)}+\lambda(1-\pi1-\pi2)
\end{split}
\end{equation}
We then differentiate each of the parameters in $\theta=(\mu_j,\sigma_{j}^2,\pi_j,\beta_j)$ and equating to zero, the parameter estimates can be maximized on $\theta$. This is illustrated in (\ref{eq:differentiation})
\begin{equation}\label{eq:differentiation}
\begin{split}
\dfrac{\partial\ln{L(\theta)}}{\partial\theta}=\mathlarger{\sum}_{t=1}^{T}\dfrac{1}{f(y_t;\theta)}\times\dfrac{\partial{f(y_t;\theta)}}{\partial\theta}
\end{split}
\end{equation}
The MS-AR model parameters can be estimated with the equations below:
\begin{equation}\label{eq:mean}
\begin{split}
\mu_j=\dfrac{\mathlarger{\sum}_{t=1}^{T}\,\,y_{t}\times Pr(s_t=j|y_t;\theta)}{\mathlarger{\sum}_{t=1}^{T}\,\,Pr(s_t=j|y_t;\theta)}
\end{split}
\end{equation}
\begin{equation}\label{eq:variance}
\begin{split}
\sigma_{j}^2=\dfrac{\mathlarger{\sum}_{t=1}^{T}\,\,(y_{t}-\mu_j)^{2}\times Pr(s_t=j|y_t;\theta)}{\mathlarger{\sum}_{t=1}^{T}\,\,Pr(s_t=j|y_t;\theta)}
\end{split}
\end{equation}
\begin{equation}\label{eq:prob}
\begin{split}
\pi_j=\dfrac{1}{T}\mathlarger{\sum}_{t=1}^{T}\,\,Pr(s_t=j|y_t;\theta)
\end{split}
\end{equation}
\begin{equation}\label{eq:coefficient}
\begin{split}
\beta_j=\dfrac{\mathlarger{\sum}_{t=1}^{T}\,\,\mathlarger{\sum}_{j=1}^{K-1}\,\,(y_{t}-\mu_j)\times Pr(s_t=j|\psi_{T};\theta)}{\mathlarger{\sum}_{t=1}^{T}\,\,\mathlarger{\sum}_{j=1}^{K-2}\,\,Pr(s_t=j|\psi_{T};\theta)}
\end{split}
\end{equation}
The expected duration of the regimes is then estimated with the equation below:
\begin{equation}\label{eq:expected_duration}
\begin{split}
E(D) = \dfrac{1}{1-p_{jj}}
\end{split}
\end{equation}

\section{Data Description}
\begin{figure}[H]
	\centering
	\includegraphics[height=05cm,width=14cm]{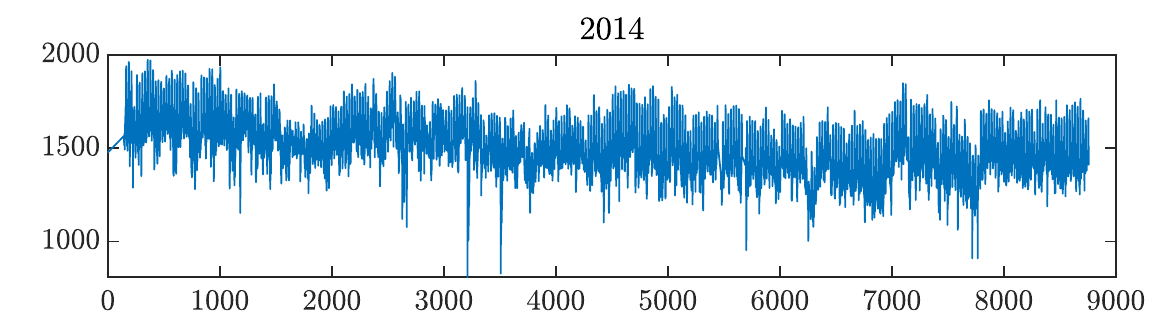}
	\caption{Plot of Hourly Electricity Consumption Time Series Data}
	\label{fig:full}
\end{figure}

\begin{figure}[H]
	\centering
	\includegraphics[height=5cm,width=14cm]{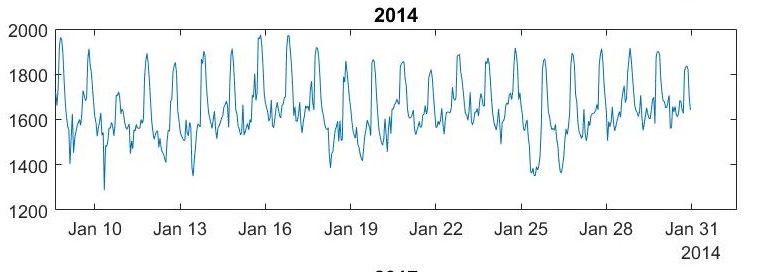}
	\caption{Illustration of electricity consumption during time of the day over a one month period}
	\label{fig:one_month}
\end{figure}

\noindent 
The electricity consumption data was obtained from the Ghana Grid Company (GRIDCO) limited. The data was analyzed from January 1, 2014, to December 31, 2014. The data obtained is the hourly power consumed by all installations linked to the distribution network. This study period is the period where Ghana was in a power crisis which was popularly referred to as ``dumsor'' period. 
\\
Multilevel seasonality is an essential characteristic of electricity consumption; the consumption pattern varies by month, day of the week, and hour of the day. These seasonalities are highly predictable and can be deterministically modelled. Figure \ref{fig:full} shows the time series plot of the hourly electricity consumption data for the period under study. The plot exhibits seasonal movement. Generally, it can be seen that the electricity consumption during the power crisis is mean-reverting, that is, it reverts back to its mean, which is about 1488 MW. Figure \ref{fig:one_month} shows changes in the consumption of electricity for a period of one month. Clearly, there are two distinct periods that can be characterized by this plot--high and low consumption periods. This can be attributed to the fact that electricity consumption during the power crisis period was lower in the early hours of the day and higher in the latter hours of the day. Each of these two periods is characterized by distinct behavioral tendencies that are inherently opposed to one another. The length of the phases that occur between the two, however, is unknown. The MS-AR approach can be used to model the electricity consumption data based on the above-mentioned conditions. The low consumption state is determined to be state one (1), while state two (2) is designated as the high consumption state. 
\\
The descriptive statistics are shown in Table \ref{table:discriptive}. The skewness values are observed to be above $0$ which indicates the consumption series is positively skewed and, as such, not normally distributed. The kurtosis is 0.2580, which implies that the distribution of the series is leptokurtic. The standard deviation value clearly shows that there is high volatility in the amount of electricity consumed during the electricity crisis period. 

\begin{table}[H]
	\caption{Descriptive Statistics of hourly electricity consumption for 2014}
	\centering
	\resizebox{\columnwidth}{!}{%
		\begin{threeparttable}
			\begin{tabular}{>{\bfseries}c*{17}{c}} 
				\toprule
				& Minimum  & Maximum & Mean & Standard Deviation & Skewness & Kurtosis 
				\\ \hline
				2014 & 809 & 1974 & 1488.0913 & 142.7847 & 0.1885 & 0.2580  \\
				\bottomrule
			\end{tabular}
			\begin{tablenotes}[para,flushleft]
				\centering{}	
			\end{tablenotes}
		\end{threeparttable}	
	}
	\label{table:discriptive}
\end{table}

\begin{figure}[H]
	\centering
	\includegraphics[height=5cm,width=14cm]{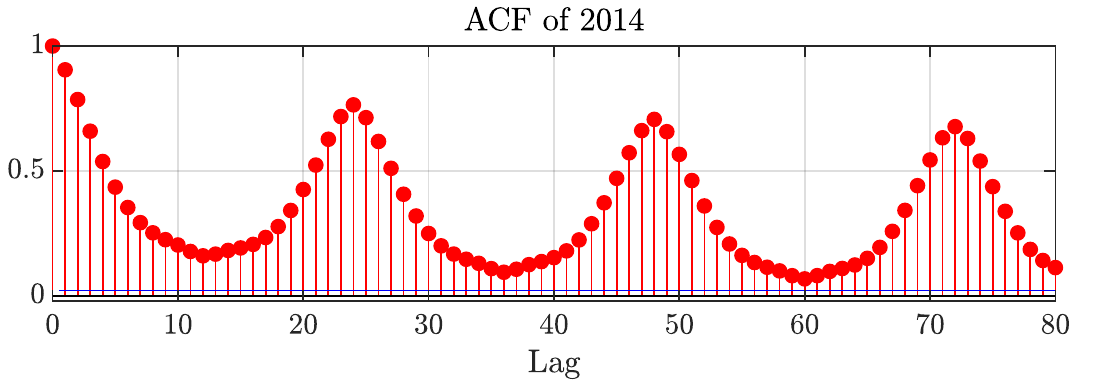}
	\caption{ACF Plot of Time Series Data 2014}
	\label{fig:acf_all}
\end{figure}

\noindent
The ACF plots in Figure (\ref{fig:acf_all}) clearly show a daily seasonal trend with peaks at 24 hours, 48 hours, 72 hours, and so on. Consider the equation below:
\begin{equation*}
L(t) = \tilde{L}_t + C_t
\label{deseasonalized_data}
\end{equation*}
where $L(t)$ is the hourly electricity consumption, $\tilde{L}_t$ is the deseasonalized electricty consumption and $C_t$ is detrministic seasonal component for $t = 0 \mbox{hour},1 \mbox{hour}, 2 \mbox{hours}, 3 \mbox{hours}, \cdots$. From Equation (\ref{deseasonalized_data}), the electricity consumption data can be deseasonalized.

\subsection{Test for Stationarity of Deseasonalized Data}
\noindent 
A stationarity test was done on the data to ascertain if a unit root exists or not. The ADF and PP tests are the methodologies used in this case. The results are presented in Table \ref{table:ADF} and \ref{table:PP} respectively. The ADF test for the three years covered by the study is summarized in Table \ref{table:ADF}. Several exogenous factors (without constant, with constant, and with constant and trend) are used in this test. The test results indicated that, regardless of the exogenous factors utilized, the series for the various years is stationary since the absolute value of the test statistics of all the three years' test results is greater than the absolute of the corresponding critical value. The PP unit root test result is shown in Table \ref{table:PP}. The series is stationary, as evidenced by the test results for all three years.

\begin{table}[H]
	\caption{ADF Test Results of hourly electricity consumption for 2011, 2014, 2017}
	\centering
	\begin{threeparttable}
		\begin{tabular}{>{\bfseries}l*{17}{c}} 
			\toprule
			Year & Test Statistics & 5\% Critical Value
			\\ \hline
			Without Constant\\
			2014	&-2.2313	&-1.95\\
			With Constant  \\
			2014	&-24.3513	&-2.86\\
			With Constant and Trend	\\	
			2014	&-27.8649	&-3.41\\
			\bottomrule
		\end{tabular}
		\begin{tablenotes}[para,flushleft]
			\centering{}	
		\end{tablenotes}
	\end{threeparttable}	
	\label{table:ADF}
\end{table}

\begin{table}[H]
	\caption{PP Test Results of hourly electricity consumption for 2011, 2014, 2017}
	\centering
	\begin{threeparttable}
		\begin{tabular}{>{\bfseries}l*{17}{c}} 
			\toprule
			Year & Test Statistics & 5\% Critical Value
			\\ \hline
			With Constant\\
			2014 	&-20.8710	&-2.862418\\
			With Trend \\
			2014 	&-23.6494	&-3.413069\\
			\bottomrule
		\end{tabular}
		\begin{tablenotes}[para,flushleft]
			\centering{}	
		\end{tablenotes}
	\end{threeparttable}	
	\label{table:PP}
\end{table}

\subsection{Correlogram}
\begin{figure}[H]
	\centering
	\includegraphics[height=05cm,width=14cm]{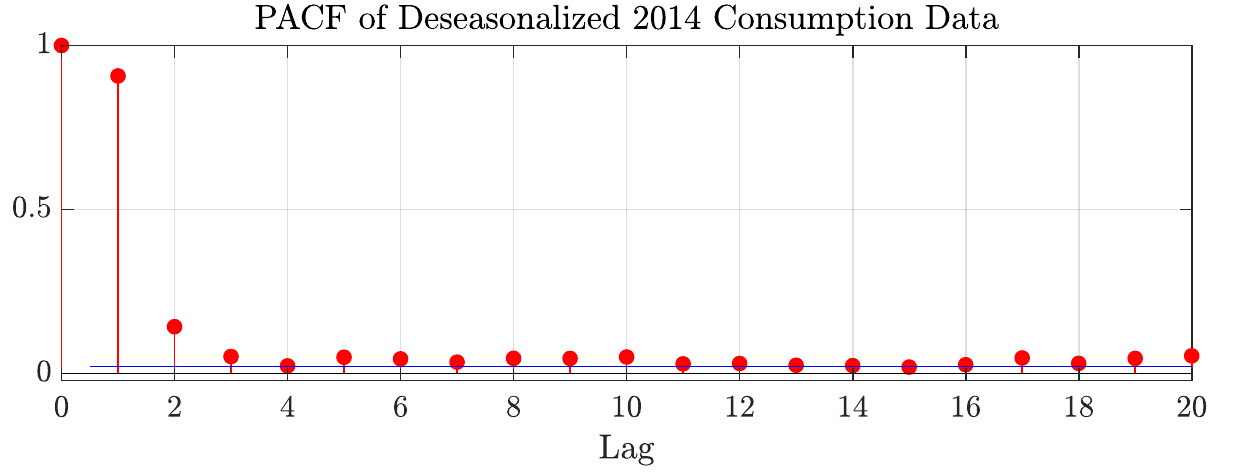}
	\caption{ACF and PACF plots of Deseasonalized 2011 Data}
	\label{fig:pacf_all}
\end{figure}

\noindent 
The number of significant autoregressive terms needed to estimate the parameters, including the transitional probabilities and autoregressive coefficients, is obtained from the partial autocorrelation function (PACF) and autocorrelation function (ACF) plots. To correctly model the electricity consumption data sets, it is essential to obtain a lag order that will result in the best autoregressive model. The PACF plot is used to establish the optimal order of the AR process since it eliminates differences caused by prior lags and keeps only the pertinent features. Utilizing the PACF also guarantees that only the necessary features or lags are kept, which is another benefit. PACF plots of the deseasonalized time series for 2011, 2014, and 2017 are shown in Figure \ref{fig:pacf_all}. For the period under consideration, lags 1, 2, 3, and 4 are statistically significant enough to be incorporated into the model.

\section{Results and Analysis}
\subsection{Markov Switching Autoregressive Model Parameter Estimation}
\begin{figure}[H]
	\centering
	\includegraphics[height=08cm,width=16cm]{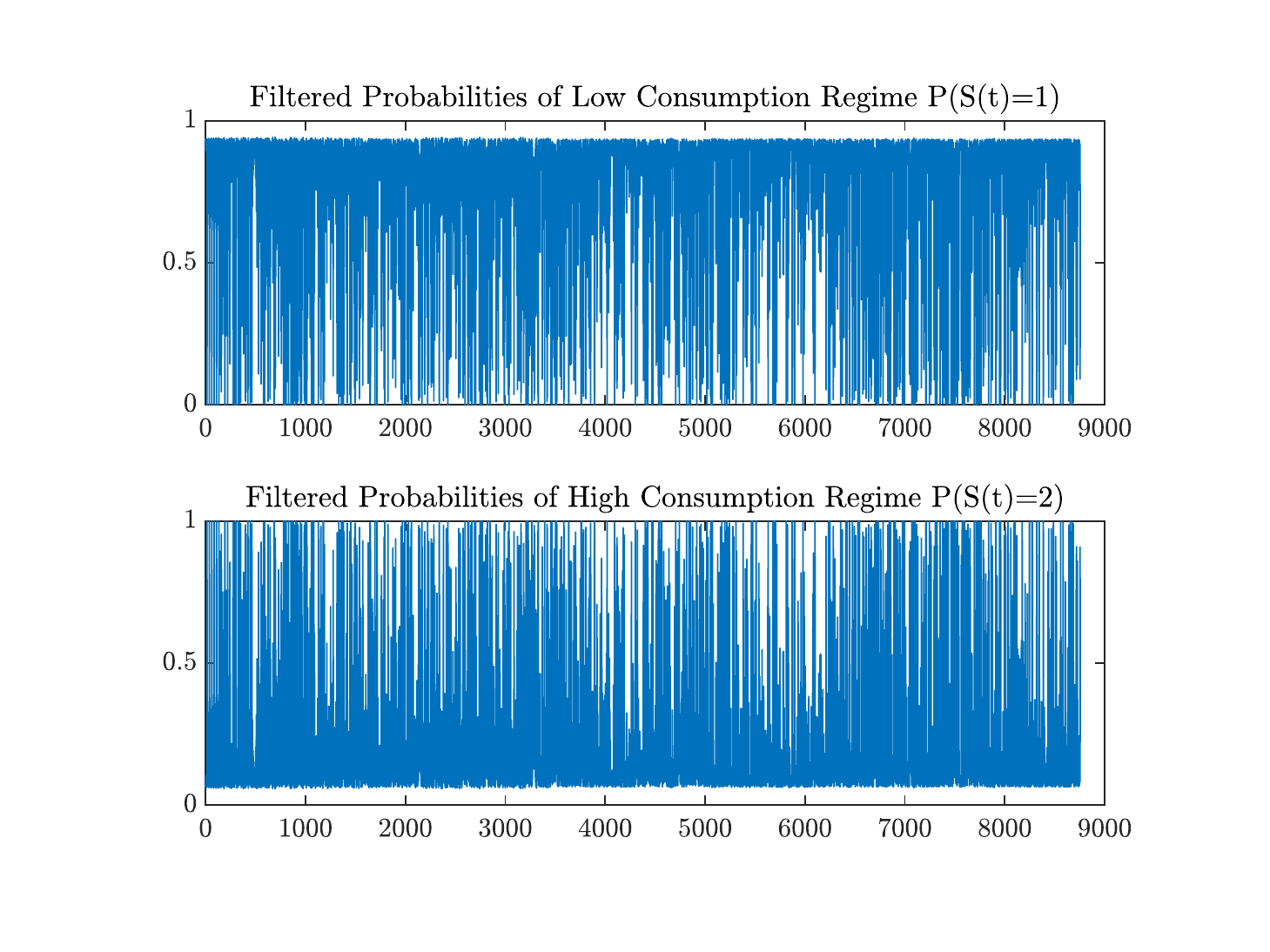}
	\caption{Filtered Probability for 2014 Hourly Electricity Consumption Data}
	\label{fig:filteredProb2014}
\end{figure}

\begin{figure}[H]
	\centering
	\includegraphics[scale=0.9]{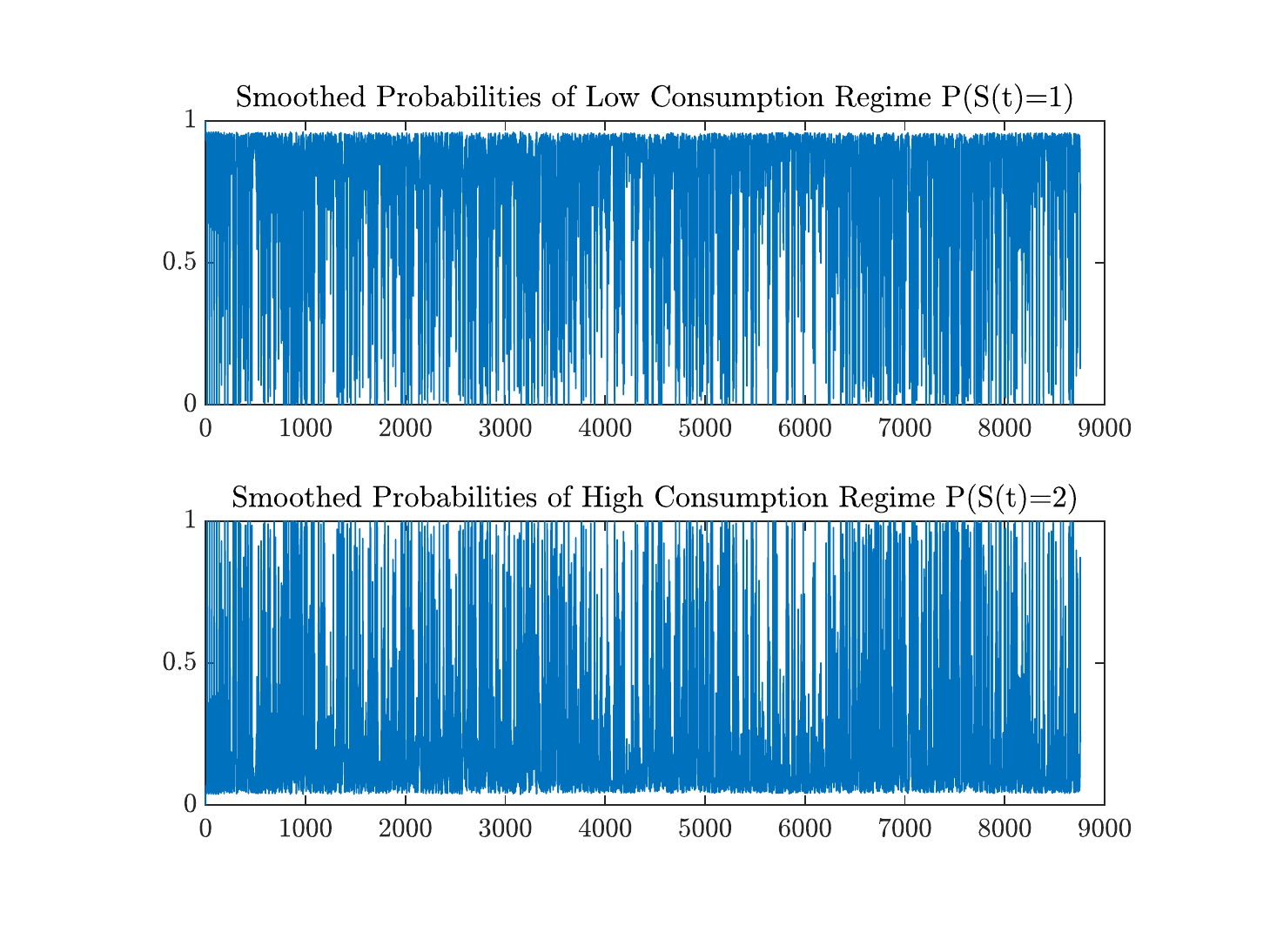}
	\caption{Smoothed Probability for 2014 Hourly Electricity Consumption Data}
	\label{fig:smoothedProb2014}
\end{figure}

\noindent 
From Figure (\ref{fig:pacf_all}), the PACF suggested we consider an AR(1), AR(2), AR(3) and AR(4) for our MS(2)-AR(p). model.
\\
To select the optimum model that describes the phenomenon, lag values 1,2,3,4 of the MS(2)-AR(p) models are analyzed and the AIC, BIC, HQC and log-likelihood values were calculated. The model with the greatest log-likelihood value but least AIC, BIC, and HQC values performs better than the others. Using the Log-likelihood, AIC, BIC, and HQC values, the best model obtained from Table (\ref{table:2014_models}) was MS(2)-AR(4).
\begin{table}[H]
	\caption{Models for 2014 Hourly Electricity Consumption Data}
	\centering
	\resizebox{\columnwidth}{!}{%
		\begin{threeparttable}
			\begin{tabular}{>{\bfseries}c*{17}{c}} 
				\toprule
				Model & AIC & BIC & HQC & Log-likelihood
				\\ \hline
				MS(2)-AR(1) &90520.77 	&90585.39 	&90530.43 	&-45256.39\\
				MS(2)-AR(2) &90413.28 	&90510.21 	&90427.75 	&-45200.64\\
				MS(2)-AR(3) &90391.16 	&90520.40 	&90410.45 	&-45187.58\\
				MS(2)-AR(4) &90381.16   &90542.71   &90405.28   &-45180.58\\
				\bottomrule
			\end{tabular}
			\begin{tablenotes}[para,flushleft]
				\centering{}	
			\end{tablenotes}
		\end{threeparttable}	
	}
	\label{table:2014_models}
\end{table}
\noindent
To ascertain the efficiency of the MS(2)-AR(4) model, it is compared with four (4) single-state models: AR(1), AR(2), AR(3), and AR(4). AIC, BIC, HQC, and log-likelihood are used to select the best model. From Table \ref{table:Comparison_MS_AR}, the AIC, BIC, and HQC values of the MS(2)-AR(4) model are the lowest, and the log-likelihood value of MS(2)-AR(4) is the largest. These indicate the superiority of the constructed MS(2)-AR(4) model in capturing the dynamics of electricity consumption through time and at different regimes. 
\begin{table}[H]
	\caption{Comparison of single-state regime models and 2-state regime Model for Hourly Electricity Consumption}
	\centering
	\begin{threeparttable}
		\begin{tabular}{>{\bfseries}c*{17}{c}} 
			\toprule
			Model & AIC & BIC & HQC & Log-likelihood
			\\ \hline
			AR(1) &92472.54 &92493.78 	&92479.78 	&-46233.27\\
			AR(2) &92293.95 &92322.26 	&92303.59	&-46142.97\\
			AR(3) &92272.28 &92307.67 	&92284.34 	&-46131.14\\
			AR(4) &92269.50 &92311.97 	&92283.97 	&-46128.75\\
			MS(2)-AR(4) & 90381.16 & 90542.71 & 90405.28   & -45180.58\\
			\bottomrule
		\end{tabular}
		\begin{tablenotes}[para,flushleft]
			\centering{}	
		\end{tablenotes}
	\end{threeparttable}	
	\label{table:Comparison_MS_AR}
\end{table}
\noindent
From the parameter estimation procedure using expectation maximization, the estimates of the model specification (MS(2)-AR(4)) of the transition probabilities and autoregressive coefficients are computed. The Table (\ref{table:2014_mp}) summarizes the estimated parameters. $\mu$ is the estimated regime constant which depends on $s_t$ and the $\beta_i$ parameters are the autoregressive coefficients for the regimes.
\begin{table}[H]
	\caption{Estimates of 2014 (during DumSor) Hourly Electricity Consumption Regime Parameters}
	\centering
	\resizebox{\columnwidth}{!}{%
		\begin{threeparttable}
			\begin{tabular}{>{\bfseries}c*{17}{c}} 
				\toprule
				Regime & Parameter & Coefficient & Std. Error & t-Statistic & p-value
				\\ \hline
				1   & $\mu$        &72.9800    &6.4039     &11.3962    &<0.0001\\
				&$\beta_1$    &0.8705     &0.0143     &60.8741 	&<0.0001\\
				&$\beta_2$    &0.0751     &0.0156     &4.8141     &<0.0001\\
				&$\beta_3$    &-0.0106    &0.0136     &-0.7794    &0.4357\\
				&$\beta_4$    &0.0171     &0.0105     &1.6286     &0.1034\\
				\\
				2   &$\mu$        &211.3742   &27.5262    &7.6790     &<0.0001\\
				&$\beta_1$    &0.6267     &0.0303     &20.6832 	&<0.0001\\
				&$\beta_2$    &0.0827     &0.0387     &2.1370     &0.03260\\
				&$\beta_3$    &0.1040     &0.0439     &2.3690     &0.01784\\
				&$\beta_4$    &0.0389     &0.0376     &1.0346     &0.30086\\
				\bottomrule
			\end{tabular}
			\begin{tablenotes}[para,flushleft]
				\centering{}	
			\end{tablenotes}
		\end{threeparttable}	
	}
	\label{table:2014_mp}
\end{table}
\noindent 
The MS(2)-AR(4) fitted model for 2014 Hourly Electricity Consumption can therefore be 
specified as;
\begin{equation}\label{eq:MS2AR42014}
y_t = \left\{
\begin{array}{lr}
0.8705(y_{t-1}-\mu_{s_{t-1}})+0.0751(y_{t-2}-\mu_{s_{t-2}})\\-0.0106(y_{t-3}-\mu_{s_{t-3}})+0.0171(y_{t-4}-\mu_{s_{t-4}}), & s_t = 1\\
0.6267(y_{t-1}-\mu_{s_{t-1}})+0.0827(y_{t-2}-\mu_{s_{t-2}})\\+0.1040(y_{t-3}-\mu_{s_{t-3}})+0.0389(y_{t-4}-\mu_{s_{t-4}}), & s_t = 2\\
\end{array}
\right.
\end{equation}
\\
The expected duration of the two regime with the transitional, smoothed, and filtered probabilities are estimated using the fitted model. The transitional probabilities for the MS(2)-AR(4) model is illustrated as a matrix bellow:
\begin{equation}\label{eq:MS2AR42014TP}
P = \left(
\begin{array}{lr}
P_{11} & P_{12}\\
P_{21} & P_{22}\\
\end{array}
\right)
= \left(
\begin{array}{lr}
0.8714 & 0.1286\\
0.4416 & 0.5584\\
\end{array}
\right)
\end{equation}
which satisfies Equation (\ref{eq:TPS}). From Matrix (\ref{eq:MS2AR42014TP}), the likelihood of the consumption of electricity remaining low during the next hour given that it was low in the previous hour is $P_{11} = 0.8714$ and the likelihood of the consumption of electricity been high during the next hour given that it was high in the previous hour is $P_{12} = 0.8714$. With a probability of $P_{22} = 0.5584$, the consumption of electricity will remain high during the next hour given that it was high in the previous hour and with a probability of $P_{21} = 0.4416$ the consumption will move to the low consumption regime.\\
The expected duration of the low electricity consumption regime in 2014 is at least 7.8 hours, while the high electricity consumption regime is expected to last at least 2.3 hours every day.\\
Figures (\ref{fig:filteredProb2014}) and (\ref{fig:smoothedProb2014}) illustrate the plots for the filtered and smoothed probability, respectively. They approximate the odds of remaining in the same regime to one or zero for a given regime. During a low consumption regime, the likelihood of being in state 1, $P(S(t)=1)$ is approximately one. This confirms $P_{11}=0.8714$ in Matrix (\ref{eq:MS2AR42014TP}).
In contrast, the likelihood of being in state 2, $P(S(t)=2)$ is approximately zero during a low consumption regime.

\begin{figure}[H]
	\centering
	\includegraphics[height=09cm,width=16cm]{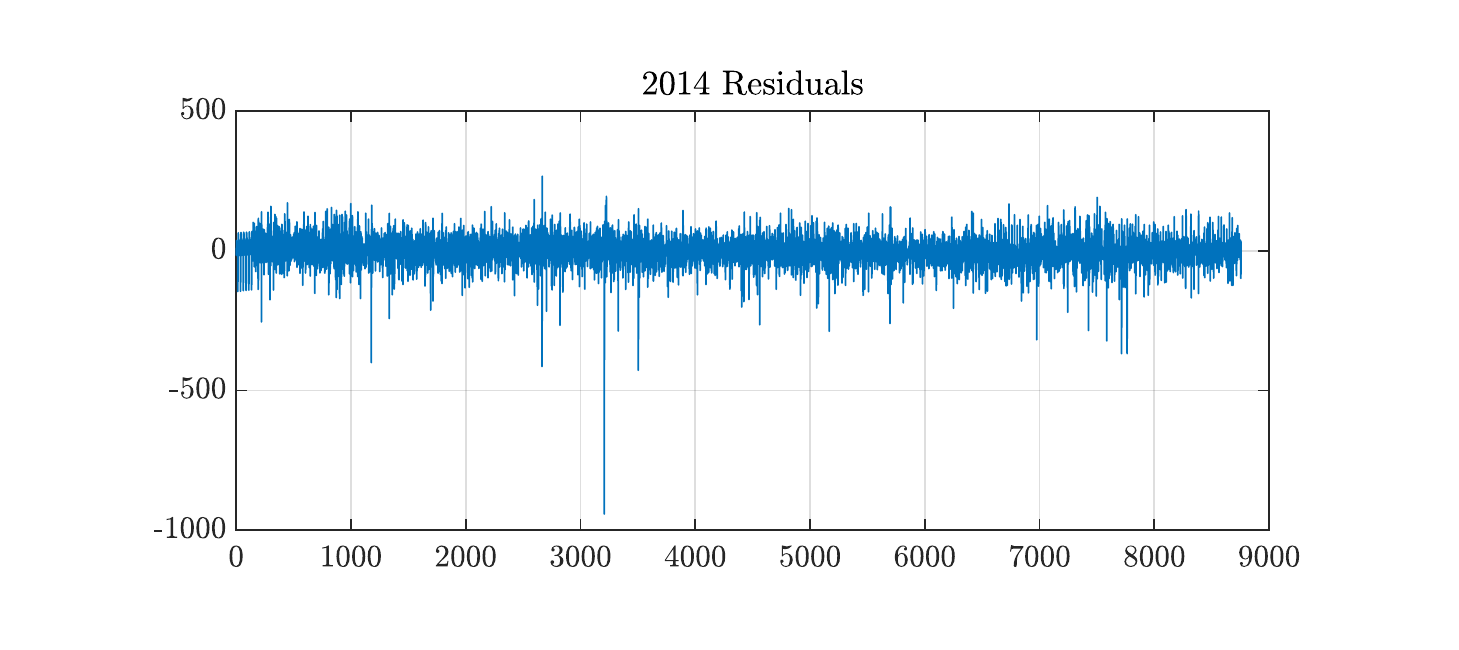}
	\caption{Residual Plot for 2014}
	\label{fig:residuals2014}
\end{figure}

\begin{figure}[H]
	\centering
	\includegraphics[height=09cm,width=16cm]{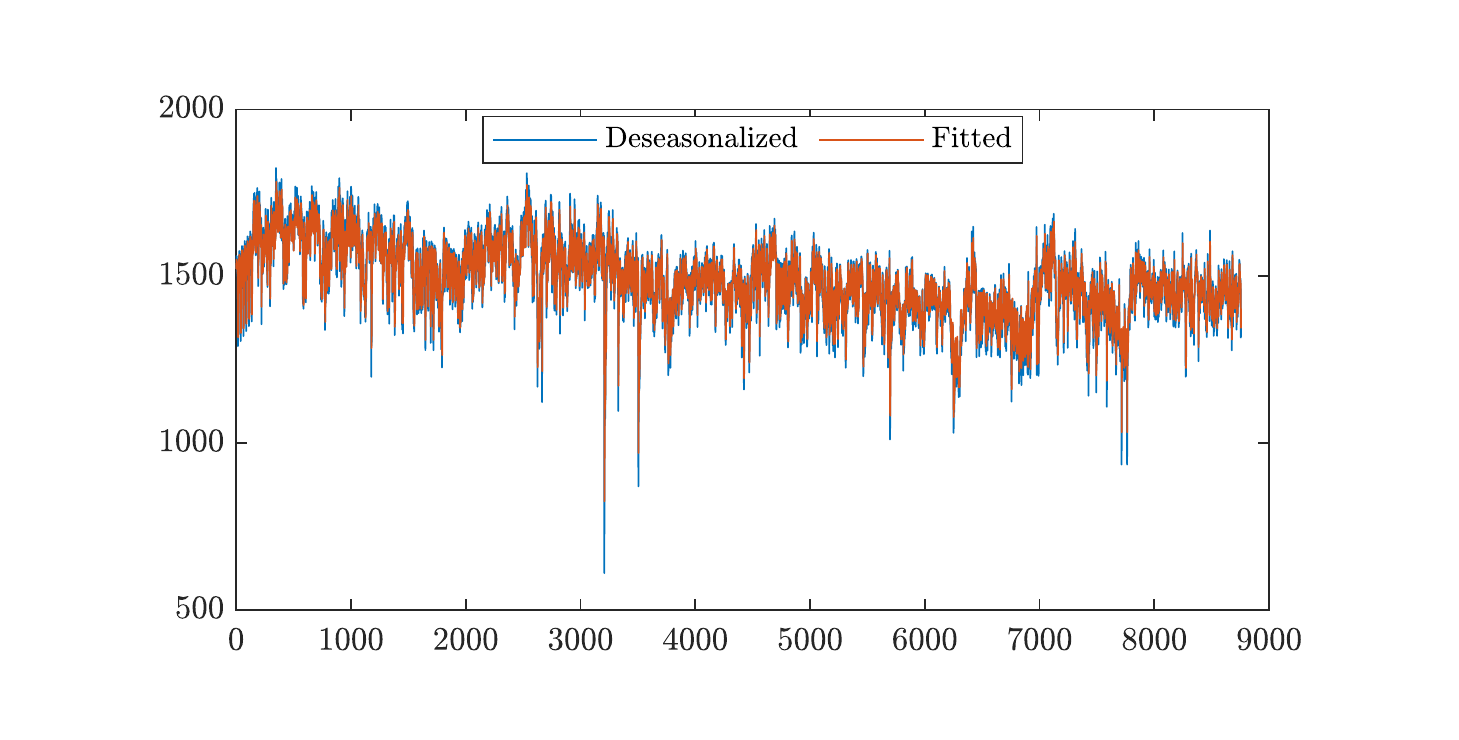}
	\caption{Deseasonalized and fitted plot of 2014}
	\label{fig:actualfitted2014}
\end{figure}

\noindent
To test the validity of the fitted model, we examined the residuals of the model with the Durbin-Watson test. The MS(2)-AR(4) has a Durbin-Watson test statistic of 2.00223, indicating uncorrelated residuals, which implies that the error terms' variances are constant. Figure (\ref{fig:residuals2014}) shows the residuals plot.
\\
Data was simulated using the estimated parameter values from the MS(2)-AR(4) model. The simulated data was fitted to the original deaseasonalized series. The deseasonalized series, as well as the fitted MS(2)-AR(4) model, are depicted in Figure (\ref{fig:actualfitted2014}). The figure shows that the proposed model is capable of effectively capturing the index's behavior over time. The suggested model can identify peaks and significant fluctuations in the electricity consumption pattern in 2014. Similarly, the model can identify distinct regime swings linked with electricity consumption.

\subsection{Discussions of Findings}
\noindent
According to the study's results, the expected duration for low and high electricity consumption periods is 7.8 and 2.3 hours, respectively, during periods of power crisis. Transitional periods also exist between these low and high consumption regimes. These are transitions from a low consumption regime to a high consumption regime and from a high consumption regime to a low consumption regime. For the year under consideration, the transition from a low to high electricity consumption regime and from a high to a low consumption regime is projected to take at least 1.1 and 1.8 hours, respectively. It is clear that electricity consumption in Ghana is low on average, with an erratic pattern. As seen in Figure (\ref{fig:full}), It also swings and varies over time, with some extremely low periods of consumption possibly due to the effects of climate change. This is reinforced by \cite{avordeh2021quantitative}, who concludes that under various scenarios of climate change, temperature rises would impact household electricity consumption significantly.

\section{Conclusion and Policy Recommendation}
\noindent 
From the study, a two-state MS(2)-AR(4) model best captures Ghana's electrical consumption pattern during the period of power crisis. According to the analysis, the likelihood of maintaining a low electricity consumption regime in 2014 is estimated to be 87\%. During the power crisis, the expected duration of a low electricity consumption regime is 7.8 hours daily. Over the course of the entire research period, the high electricity consumption regime is expected to last 2.3 hours daily.The study calculated the probability of transitioning from one regime to another as well as the predicted duration for each regime. This information should be used by electricity distributors to measure the amount to be distributed on an hourly basis during periods of a power crisis.

%

\section*{Declarations}
\subsection*{Funding}
\noindent
This work did not receive any funding.  

\subsection*{Competing Interests}
\noindent
There are no competing interests with the publication of the present manuscript.
\subsection*{Availability of data and materials}
\noindent
Data are available from the corresponding author upon request.

\bibliographystyle{model1-num-names}
\bibliography{sample.bib}

%
%
%
%

\end{document}